\documentclass[aps,prl,twocolumn,letterpaper,superscriptaddress,floatfix,showpacs]{revtex4-1}

\usepackage{graphicx,amsmath,amsfonts,amssymb,bm,tabularx}
\usepackage[normalem]{ulem}
\usepackage{color,wrapfig}
\usepackage{float}
\def\be{\begin{equation}}
\def\ee{\end{equation}}

\def\betaDPsc{$\beta^{\prime \prime}$-(ET)$_2$SF$_5$CH$_2$CF$_2$SO$_3$}

\def\kETncs{$\kappa$-(ET)$_2$Cu(NCS)$_2$}

\def\t1{$T_1^{-1}$}

\def\ita{\textit{a}}
\def\itb{\textit{b}}

\usepackage[usenames,dvipsnames,svgnames,table]{xcolor}

\DeclareFontFamily{OT1}{pzc}{}
\DeclareFontShape{OT1}{pzc}{m}{it}{<-> s * [1.2] pzcmi7t}{}
\DeclareMathAlphabet{\mathpzc}{OT1}{pzc}{m}{it}

\begin{document}

\title{Microscopic study of the Fulde-Ferrell-Larkin-Ovchinnikov state in an all-organic superconductor}

\author{G. Koutroulakis}
\affiliation{Department of Physics $\&$ Astronomy, UCLA, Los Angeles, CA 90095, USA}
\author{H. K\"{u}hne}
\affiliation{Hochfeld-Magnetlabor Dresden (HLD-EMFL), Helmholtz-Zentrum Dresden-Rossendorf, D-01314 Dresden, Germany}
\author{J. A. Schlueter}
\affiliation{Materials Science Division, Argonne National Laboratory, Argonne, IL  60439 USA}
\author{J. Wosnitza}
\affiliation{Hochfeld-Magnetlabor Dresden (HLD-EMFL), Helmholtz-Zentrum Dresden-Rossendorf, D-01314 Dresden, Germany}
\affiliation{Institut f\"{u}r Festk\"{o}rperphysik, TU Dresden, D-01069 Dresden, Germany}
\author{S. E. Brown}
\affiliation{Department of Physics $\&$ Astronomy, UCLA, Los Angeles, CA 90095, USA}

\voffset=0.5cm
\begin{abstract}
Quasi-two dimensional superconductors with sufficiently weak interlayer coupling allow magnetic flux to penetrate in the form of Josephson vortices for in-plane applied magnetic fields. A consequence is the dominance of the Zeeman interaction over orbital effects. In the clean limit, the normal state is favored over superconductivity for fields greater than the paramagnetic limiting field, unless an intermediate, inhomogeneous state is stabilized. Presented here are nuclear magnetic resonance (NMR) studies of the inhomogeneous (FFLO) state for \betaDPsc. The uniform superconductivity-FFLO transition is identified at an applied field value of 9.3(0.1) T at low temperature ($T=130$ mK), and evidence for a possible second transition between inhomogeneous states at $\sim11$ T is presented. The spin polarization distribution inferred from the NMR absorption spectrum compares favorably to a single-Q modulation of the superconducting order parameter.


\end{abstract}

\maketitle

Superconducting charge transfer salts based on the organic donor molecule bisethylenedithio-tetrathiafulvalene (ET) are associated with very anisotropic electronic properties, due to the layered arrangements of ET molecules separated by anionic spacers \cite{Ishiguro:1998,Singleton:2002,Lebed:2008}. The details of the planar arrangement provide a classification criterion, allowing for meaningful comparison of physical properties between what are essentially isomorphic compounds. The most familiar polymorph is probably $\kappa$-(ET)$_2$$X$, where the designator $\kappa$ is associated with a parquet-like arrangement of dimerized ET molecules, and the anions $X$ are polymerized inorganic ligands such as Cu[N(CN)$_2$]Br and Cu(NCS)$_2$. In comparing the properties amongst those salts, there is clear evidence for the importance of correlations, which are enhanced because of the weak interlayer coupling. This feature, along with the observed long mean free paths \cite{Wosnitza:1996,Singleton:2000a}, make them ideal materials for investigating both the possibility for, and the nature of, field-induced superconducting (SC) phases, stabilized near to and beyond the paramagnetic limiting field $B_P$ \cite{Singleton:2000,Lortz:2007,Wright:2011,Bergk:2011,Agosta:2012,Beyer:2012,Mayaffre:2014}.

In such cases, a first-order phase transition from SC to normal states was predicted, driven by the Zeeman interaction \cite{Clogston:1962}. However, under fairly restrictive circumstances, various intermediate phases were also suggested. The original proposals, by Fulde and Ferrell \cite{FF:1964} and independently by Larkin and Ovchinnikov (LO) \cite{LO:1965}, were made $\sim$50 years ago, and therein the principle mechanism and constraints were identified. Namely, electron pairs acquire a non-zero momentum $\mathbf{q}$ as a consequence of the applied field. For the LO case, the gap oscillates through zero in real-space, $\Delta(\vec{r})\sim\cos(\vec{Q}\cdot\vec{r})$. In higher than one dimension, the Fermi surface of the compromise state is only partially removed by the pairing. Then, depending on details, such as Fermi surface structure, order parameter symmetry of the low-field SC state (uSC), and field range, single or multiple momentum components are close in energy \cite{Matsuda:2007}.

Compelling evidence for FFLO physics in real materials is quite recent and restricted to just a few layered molecular superconductors, such as \kETncs\ (hereafter $\kappa$-NCS) \cite{Wright:2011,Bergk:2011,Agosta:2012,Mayaffre:2014} and $\lambda$-(BETS)$_2$FeCl$_4$ \cite{Uji:2006}. Relevant here is that they are in the clean-limit, remarkably anisotropic, and $B_P$ is accessible (albeit using the resistive magnets of the major facilities). The more familiar magnetic-field coupling to the SC state, which occurs through vortex creation and order parameter suppression in the cores, is avoided for short interlayer coherence and in-plane fields, since flux penetration occurs in the form of Josephson vortices \cite{Mansky:1994,Kirtley:1999}, and a straightforward geometric consequence is that the in-plane screening currents fall off as $1/B$ \cite{Bulaevskii:1991}. 

Most studied is $\kappa$-NCS, at the macroscopic level using transport \cite{Singleton:2000,Agosta:2012}, specific heat \cite{Lortz:2007}, and torque magnetometry probes \cite{Bergk:2011}, and microscopically with $^{13}$C NMR spectroscopy \cite{Wright:2011} and relaxation experiments \cite{Mayaffre:2014}. The uSC-FFLO boundary was distinct in both types of NMR measurements. The spectroscopy demonstrated the high-field phase was characterized by a sudden increase in both the mean spin polarization $\overline{M}_s(\vec{r})$, and a broad real-space distribution of $M_s(\vec{r})$ while remaining a bulk superconductor. The temperature dependence of the relaxation rate was interpreted as evidence for real-space gap zeroes of an LO state. In considering other systems generally, and the $\beta''$ particularly, our aims are several. Since it is just the second candidate system studied microscopically, we consider an NMR investigation central to confirmation of FFLO, since it is specifically sensitive to electron spin states. Moreover, what physics is generic to FFLO or particular to $\kappa$-NCS is an open question. Even then, the form of the modulation and how it evolves with field is undetermined. Finally, it is well-known that inhomogeneous phases are sensitive to disorder; with accessible FFLO phases, how the high-field state is impacted should have experimental consequences.

Recent specific heat studies of \betaDPsc, with $T_c=4.3$ K, support the FFLO scenario in the range of 9-10 T and above, which is accessible to standard laboratory magnets \cite{Beyer:2012}. Reported here are $^{13}$C NMR spectroscopy and relaxation measurements to 11.9 T. A phase transition is identified at $B_{s}\simeq9.3$ T, beyond which significant line-broadening is observed in what is a bulk superconductor (as it must be for FFLO). The spectrum is shown consistent with a single-Q modulation. We conclude also that there is considerable local variation in the spatial modulation of the SC gap, which we tentatively associate with the relative importance of disorder in the high-field phase and which we expect also impacts the FFLO-normal phase boundary. Finally, the possibility for a 2$^{nd}$ field-induced transition is indicated in the field-dependence of the relaxation rate.

ET molecules with $^{13}$C spin-labelled on the ET central carbons were used in the electrolytic single-crystal growth of \betaDPsc. The crystal chosen had well-defined faces, with dimensions approximately $0.8$ mm $\times 4.3$ mm $\times 0.1$ mm and mass 0.9 mg. The long direction is the \itb-axis, and the shortest dimension is interlayer. Our goal was to collect NMR spectra at low temperature and over as wide a range of calibrated fields as possible, applied precisely in-plane. For that purpose, a small coil, with dimensions appropriate for high filling factor, was mounted on a piezo-rotator (order millidegree angular steps) with its axis aligned with the long dimension of the crystal. For exploring the SC phase, the sample was cooled to $T< 1$ K in a dilution refrigerator, with the experiment placed directly into the mixing chamber for good thermalization. The NMR experiments were performed using a top-tuning tank circuit, which avoids the problem of inserting mechanically-adjusted circuit elements into the mixing chamber but allows for measurements over a wide field range. A drawback is signal loss due to attenuation in the cable.

\begin{figure}[ht]
\includegraphics[width=3.5in]{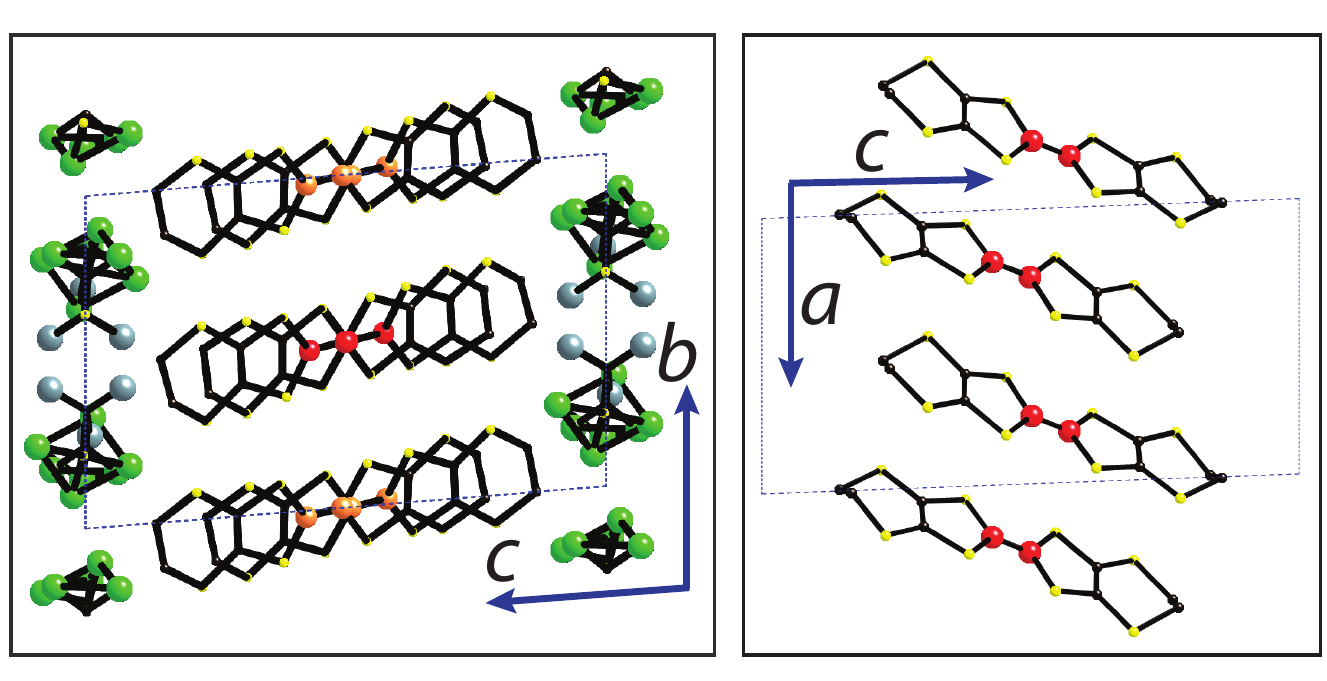}
\includegraphics[width=3.5in]{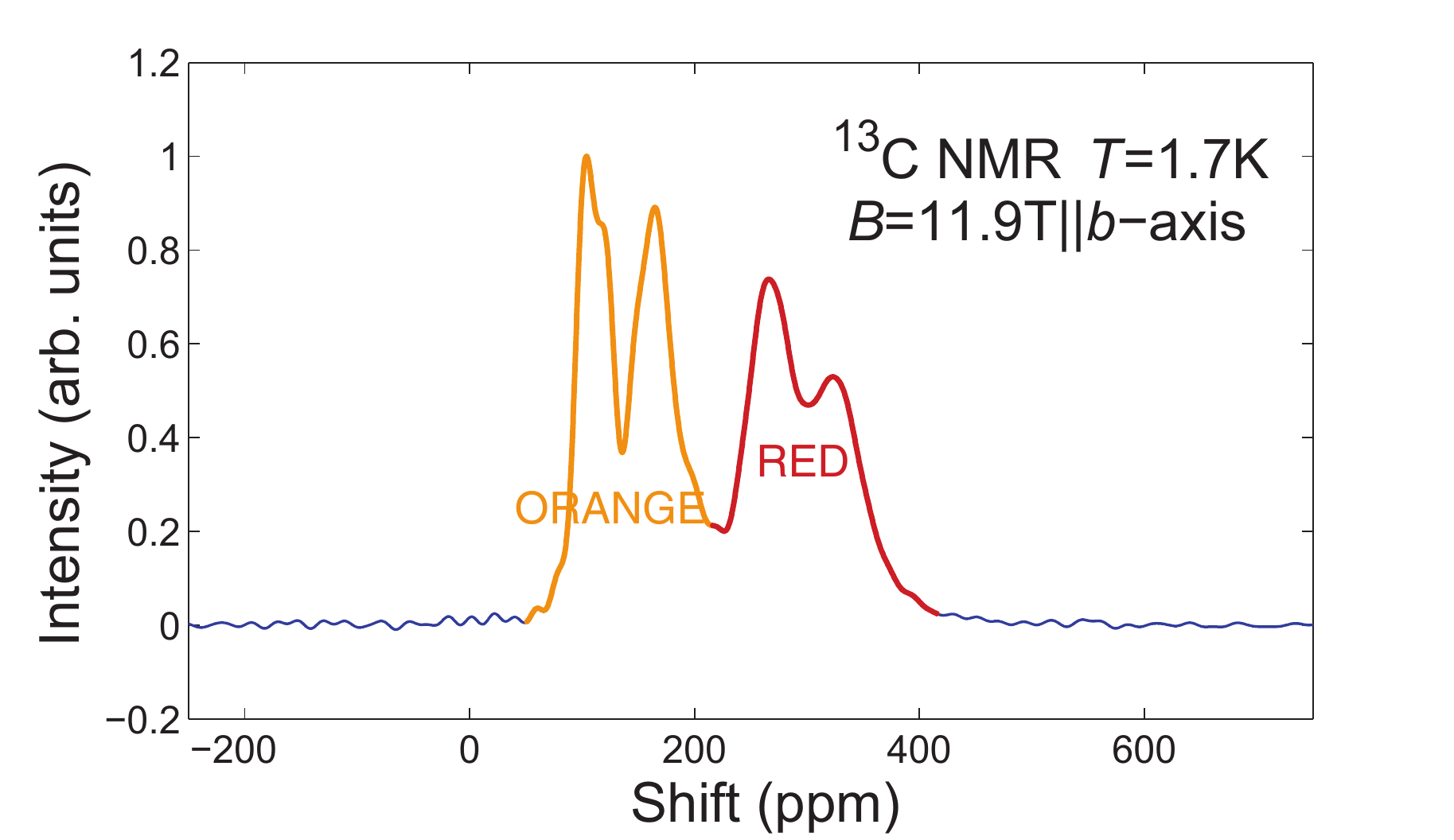}
\caption{(\textit{top}) Crystal structure, viewed along [100] (left) and [010] (right) directions. Inequivalent molecules are segregated in stacks aligned with $a$. The $^{13}$C sites are indicated by the colors orange or red, with the \textit{red} molecules sandwiched by the negatively charged SO$_3$ groups (grey spheres). (\textit{bottom}) $^{13}$C NMR spectrum, recorded at $B=11.9$ T and $T=1.7$ K, corresponding to the normal state, with the color-coding indicating the contributions from the \textit{red} and \textit{orange} sites. The applied field is directed precisely in the plane formed by the ET layers, approximately $\perp\mathbf{b}$.}
 \label{fig:TwoD}
\end{figure}

$^{13}$C is commonly introduced into ET donors for the purpose of probing the hyperfine fields, most effective are the dimer bridging sites. The other choices available here, $^1$H and $^{19}$F, are too weakly coupled. Even then, since carbon is relatively light, the measured paramagnetic shifts are small, typically of order 100-300 ppm relative to the standard reference ($^{13}$CH$_3$)$_4$Si (TMS). In Fig. \ref{fig:TwoD}(\textit{top}), the crystal structure is shown, with views along \ita\ (left) and \itb\ (right). The former emphasizes the layered structure, with ET donors configured in sheets and separated by the organic anions \cite{SuppInfo}. The labelled $^{13}$C ions are highlighted in red and orange for two crystallographically inequivalent molecules. Within the layers, the underlying morphology is aligned stacks of ET donors (into the page). In Fig. \ref{fig:TwoD}(\textit{bottom}), a $^{13}$C NMR spectrum is shown, recorded at $T=1.7$ K with magnetic field $B=11.9$ T, aligned precisely within the layers and approximately $\perp$ \itb. These conditions correspond to the normal state. In the crystal lattice, the two $^{13}$C sites of each of the two molecules are inequivalent, making for four independent sites. The contributions from the two molecules are well-resolved: the absorption from one of the two molecules appears at higher frequency than the other. The distinction arises from a substantial charge carrier imbalance between the two stacks. Our interpretation is that the ET stack sandwiched by the negatively-charged SO$_3$ ligands of the counterion has the higher hole density, and thus, the greater shift. Following convention \cite{DeSoto:1995}, we label the two sites within each molecule ``inner'' and ``outer''; the shifts are generally greater for the outer site, because it is positioned closer to the negatively-charged counterion \cite{DipolarNote}. The contributions are labelled in Fig. \ref{fig:TwoD}.

\begin{figure}[ht]
\includegraphics[width=3.5in]{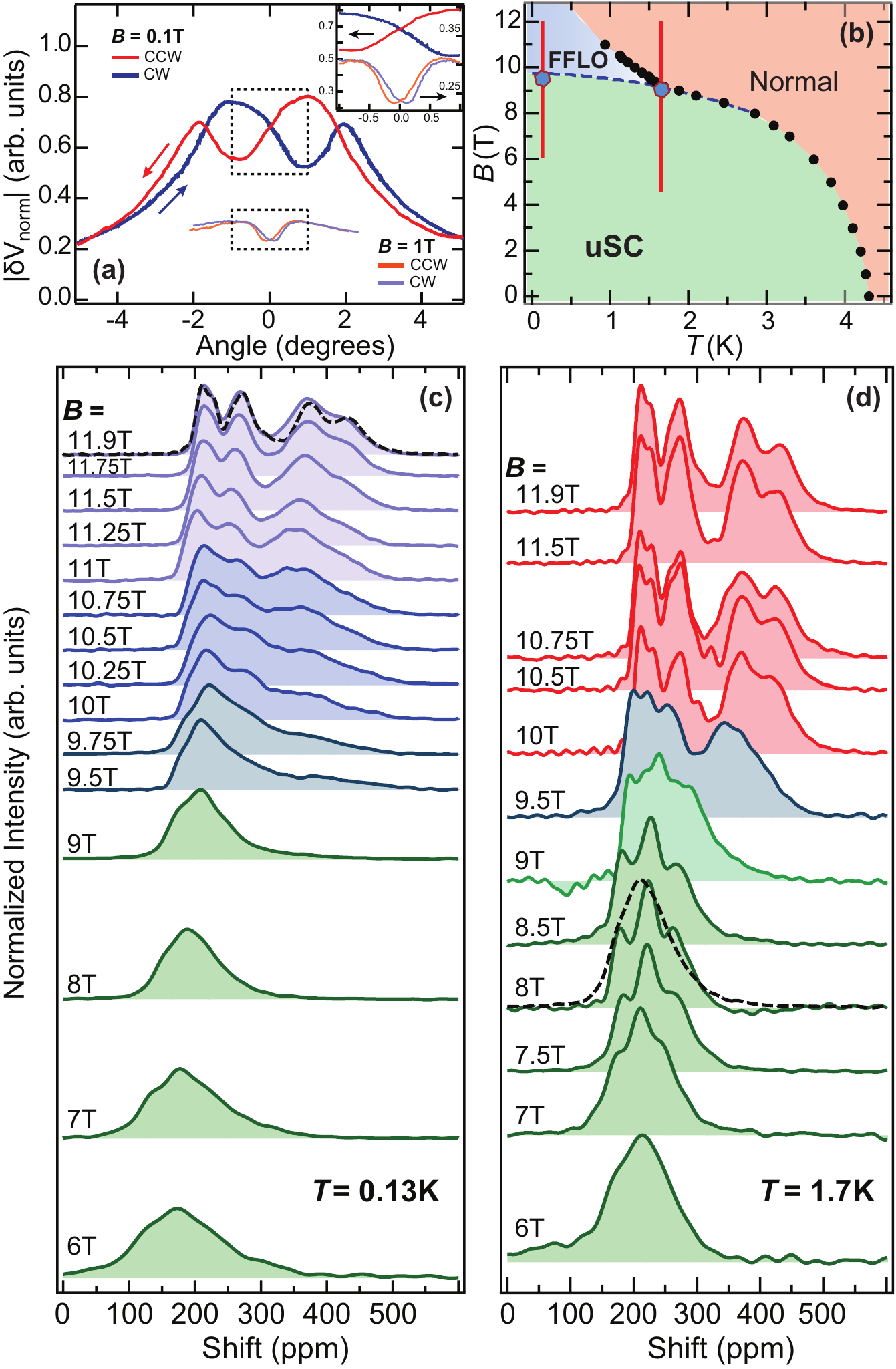}
\caption{(a) Angle dependence of the reflection signal, with the anomaly at $\theta$=0 associated with the in-plane condition. Hysteresis is reduced upon increasing the field strength. (b) Field/temperature superconductor/normal state phase boundary, as inferred from specific heat measurements (black circles) \cite{Beyer:2012}, and the NMR results reported here (blue diamonds). The vertical red bars correspond to the range of field strength for our measurements (representative spectra shown in (c) and (d)). (c) Spectral evolution with field, recorded at $T$=130 mK and (d) $T$=1.7 K. At the lower temperature, significant line-broadening is exhibited for fields greater than 9.5T, which we take to be onset of the FFLO phase. For $T$=1.7 K, only the spectrum at 9.5 T takes on a different, transitional, form from what appears at lower and higher fields. }
 \label{fig:SpectraVsB}
\end{figure}

Specific heat measurements have demonstrated the importance of in-plane fields to within less than 1$^{\circ}$ alignment \cite{Beyer:2012}. For this purpose, we relied on the angle dependence of normal-core vortex creation and dynamics, which are identifiable through measurements of the RF complex impedance (and in the high-field NMR spectrum \cite{SuppInfo}). Specifically, the circuit was tuned and matched (to $\mathit{f}$=13.4 MHz), and changes to one channel of the complex RF reflection were monitored upon varying the angle of the applied field at $T$=1.7 K. In Fig. \ref{fig:SpectraVsB}(a) is shown results for sweeps in both directions, which includes some hysteresis. The anomaly centered at $\theta$ = 0 marks the in-plane condition \cite{Shinagawa:2007}, with the field penetrating as Josephson vortices. Note that the action of the applied RF is to oscillate the field in the plane formed by the dc-field direction and the coil's symmetry axis. Therefore, when aligned, the flux penetration remains in-plane, the Josephson-vortex lattice is weakly pinned, and there is good RF penetration and signal strength despite the superconductivity.

In Fig. \ref{fig:SpectraVsB}(c),(d), the field dependence of the absorption spectrum is shown, recorded at $T$=130 mK, 1.7 K, respectively, and plotted as shift (ppm). At $T$=130 mK, the onset of the inhomogeneous electron spin polarization occurs for fields exceeding 9.3(0.1) T, marked by an increase in overall linewidth. At the higher fields, the features corresponding to the four inequivalent sites are identifiable. For clarity, the normal-state spectrum of Fig. \ref{fig:TwoD} is also shown as the black dashed trace in Fig. \ref{fig:SpectraVsB}(c). 

For comparison, the phase diagram determined by specific heat measurements (solid black circles) \cite{Beyer:2012} appears in Fig. \ref{fig:SpectraVsB}(b), where the transition between uniform SC and FFLO states identified by the NMR results reported here is denoted (blue diamonds with red borders). The vertical red bars indicate the range of fields covered by the NMR measurements, recorded at $T$=130 mK, 1.7 K.

\begin{figure}
\includegraphics[width=3.5in]{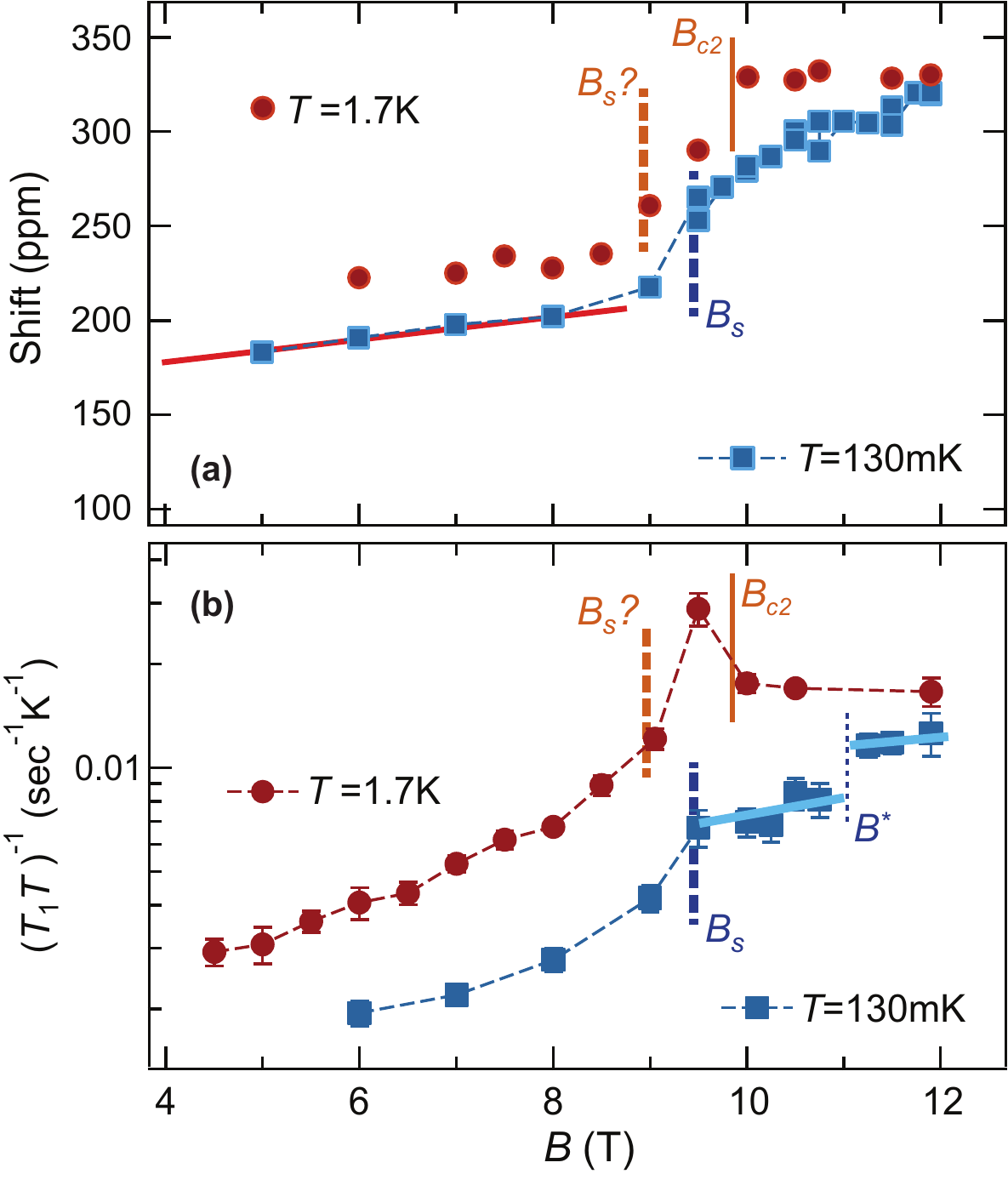}
\caption{(a) $^{13}$C NMR shift \textit{vs}. $B$. The solid red line corresponds to the Zeeman effect on a $d$-wave model for circular Fermi surface and weak coupling (see text). Its extrapolation to $B=0$ leads to a 150 ppm estimate for the orbital shift. (b) Spin-lattice relaxation rate, $T_1^{-1}(B)$. The data recorded at $T$ = 1.7 K exhibit an increase at 9.3 T greater than the normal state value, consistent with an incursion into the FFLO phase \cite{Mayaffre:2014,ToneNote}. The curves connecting the data points are guides-to-the-eye.}
 \label{fig:NMRvsB}
\end{figure}

A first step in quantifying the FFLO state can be found in the $^{13}$C first moment NMR shift, $\delta\nu(B)/\nu_0 \equiv (\overline\nu-\nu_0)/\nu_0$, with $\nu_{0}$ the reference frequency (for TMS). At low field and deep in the uSC state, only the orbital part contributes intrinsically to the shift. Whereas, the constant shift of the normal state includes the hyperfine fields associated to the spin susceptibility. Fig. \ref{fig:NMRvsB}(a) depicts the field dependence of $\delta\nu(B)/\nu_0$. The variation between the two limits is associated with the increase of the hyperfine fields, most rapidly at the onset of the inhomogeneous (FFLO) SC phase (blue squares) at $B_s\simeq9.3$ T. Complementary results for the spin-lattice relaxation rate are shown in Fig. \ref{fig:NMRvsB}(b). Since the $^{13}$C magnetization recovery varies considerably over the whole linewidth, what is plotted is the time scale associated with 63\% of the full recovery (1-1/$e$) \cite{ToneNote}.

Ideally, how the high-field SC phase (FFLO) should be interpreted depends in part on our understanding of the low-field phase (uSC) \cite{Matsuda:2007,Vorontsov:2005}, as well as on further evidence for phase transitions within the inhomogeneous phase. The increasing shift for 5-9 T, shown in Fig. \ref{fig:NMRvsB}, is expected in the case of momentum-space nodes. Consider the simple model of a $d$-wave order parameter with amplitude constrained by the weak-coupling result $2\Delta_0/k_BT_c$=4.3, on a circular Fermi surface. From the resulting Zeeman shift of quasiparticle energies, the result, in the applicable limit $\mu_BB/k_BT\gg1$, is $M_s/M_n=\mu_BB/2\Delta_{0}$, with $M_s$ ($M_n$) the magnetization of the SC (normal) state \cite{Yang:1998}. Then, $2\Delta_0/k_B\sim18$ K and $M_s/M_n\approx$4 \%/T, which corresponds to the red line in Fig. \ref{fig:NMRvsB}(a) \cite{Koutroulakis:2015}. (In this context, we note that controversy remains regarding SC states in ET-based superconductors. See, \textit{e.g.}, Refs. \cite{Lortz:2007,DeSoto:1995}).

\begin{figure}[ht]
\includegraphics[width=3.5in]{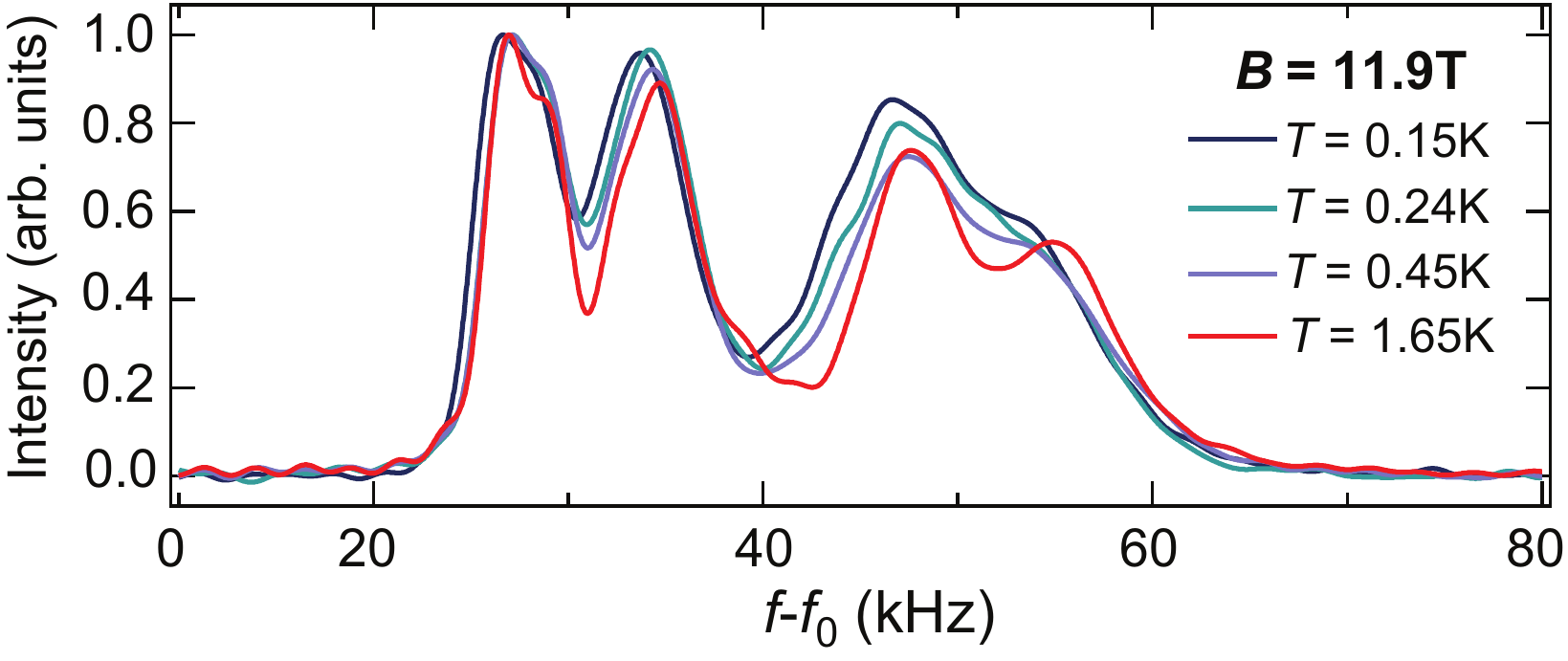}
\caption{Temperature dependence of the $^{13}$C spectrum, carried out at an applied field $B=11.9$ T. The variation is due to an increasing hyperfine field upon warming.}
\label{fig:SpecTdep}
\end{figure}

In relation to the inhomogeneous phase, the observed rapid increase of the shift for $B>B_s$ = 9.3(0.1) T  compares favorably to expectations for the onset of an FFLO state in a $d$-wave superconductor \cite{Vorontsov:2006}, and is similar to prior results from $\kappa$-NCS. In addition, here there is more than one indication for an additional phase transition at $B^*\sim$ 11 T, but with bulk superconductivity surviving to a greater applied field. From Fig. \ref{fig:NMRvsB}(b), the locations of the transitions, \textit{i.e.}, the FFLO onset at $B_s$ = 9.3 T and the possibility for another at $B^*$, have signatures in the field-dependence of $T_1^{-1}$. Also, for $B>B^*$, the spectral features are seen to narrow considerably (Fig. \ref{fig:SpectraVsB}(c)). The evidence for bulk superconductivity persisting at least to $11.9$ T at $130$ mK appears in Fig. \ref{fig:SpecTdep}, which shows the progressive increase of shift upon warming. Note that the changes are larger for the sites with greater hyperfine fields.

The structure of the FFLO phase may evolve with field, even for simple layered $s$-wave superconductors. For example, in Ref. \cite{Matsuda:2007}, the sequence of high-field phases for a two-dimensional $s$-wave superconductor is described, where the single-Q phase is destabilized in favor of double-Q structures. In the case of order parameter nodes, as for a $d$-wave superconductor, different constraints are imposed on the wavevector and, again, more than one transition is expected within the FFLO state \cite{Vorontsov:2005}. If this situation were to apply to the $\beta^{\prime \prime}$ superconductor discussed here, it is tempting to assign the lower  transition, at $B_s$, as uSC to FFLO (FFLO$_1$) and the possible upper transition, at $B^*$, as between inhomogeneous phases FFLO$_1$ and FFLO$_2$. Further study is required to confirm the FFLO$_1$ $\to$ FFLO$_2$ transition for this material.

\begin{figure}
\includegraphics[width=3.5in]{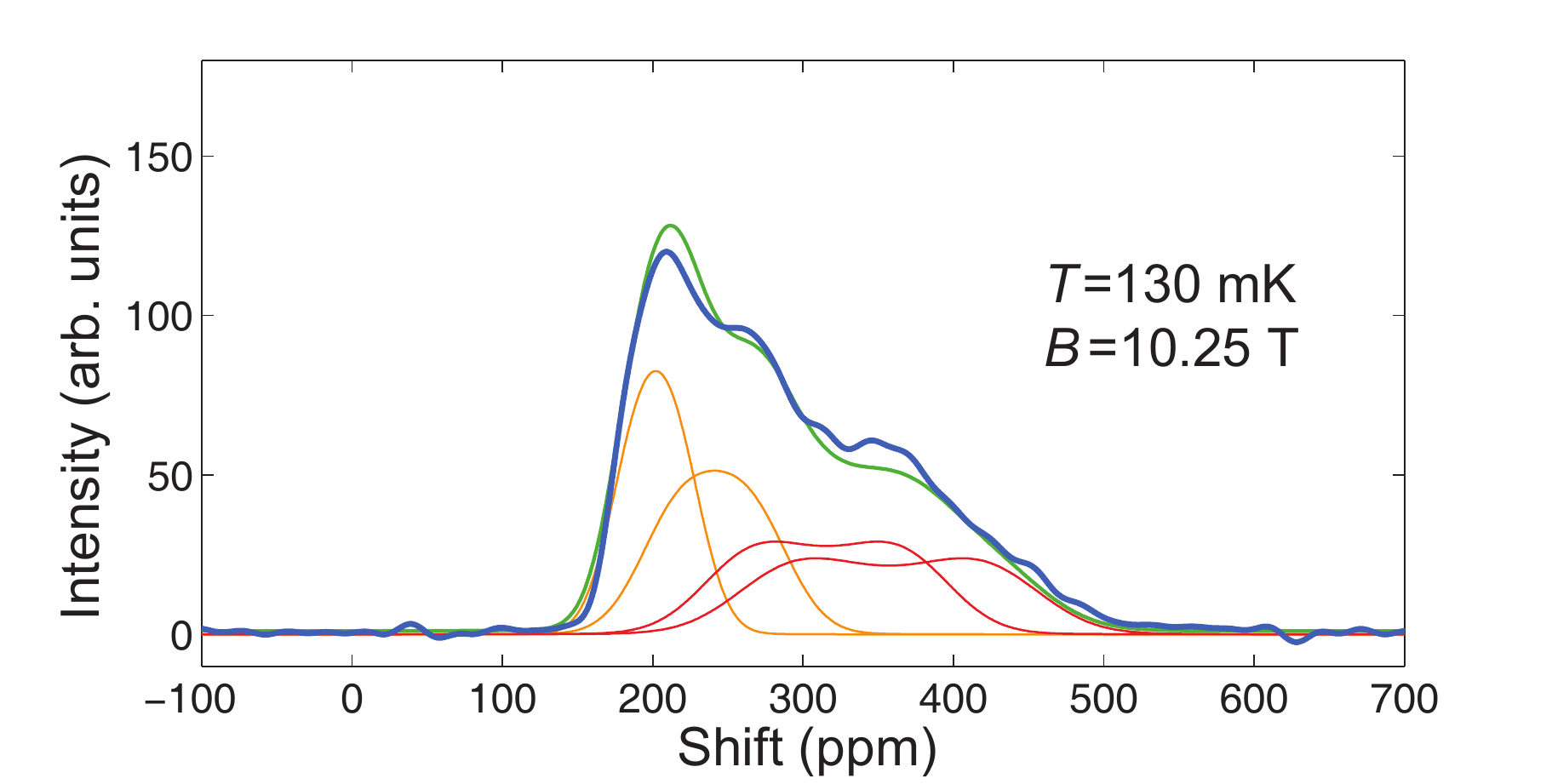}
\caption{Example of spectrum simulation in the FFLO state, compared to recorded spectrum (blue). The simulation is a sum (green) of four gaussian-broadened contributions (red, orange) arising from a single-Q sinusoidal modulation of the SC order parameter \cite{SuppInfo}.}
 \label{fig:Spec_Sim}
\end{figure}

Finally, we address the task of modeling the inhomogeneous electron spin polarization in the FFLO phase. The result is not unambiguous, since the lineshapes result from four separate contributions with equal total intensity but different hyperfine couplings and local fields. Thus, we see that while the organic conductors may be ideal for exploring superconductivity beyond the paramagnetic limit, NMR as a probe suffers from the low symmetry inherent to these compounds. Therefore, we take the simplest approach possible, and ask whether single-Q sinusoidal modulation of the SC order parameter is consistent with the data. It is, provided that the idealized lineshape is broadened, for example, by coupling to disorder. The modulation amplitude is fixed to scale with the longitudinal hyperfine coupling of each of the four contributions \cite{SuppInfo}. We justify this approximate approach as follows. At the FFLO critical field, the order parameter would vary near the zero crossing as an isolated solitonic domain wall of width of the order of the coherence length. However, a soliton lattice forms immediately once moving into the modulated phase, with lattice constant roughly given by the same length scale. At higher fields, the modulation is sinusoidal with progressively weakened amplitude. Then, the modulated part of the spin polarization varies as the modulus of the gap amplitude $|\Delta(x)|^2$. The four contributions are modeled and displayed in Fig. \ref{fig:Spec_Sim} along with their sum, for contrast to the spectrum recorded at 10.25 T.

In summary, presented here is evidence for a field-induced transition to an inhomogeneous SC phase, an FFLO state, at $B_s$ = 9.3(0.1) T and within the limits of the superconducting phase diagram of the all-organic material \betaDPsc. Moreover, a possible second phase transition is identified, between inhomogeneous phases at $\sim$ 11 T. Further study of the field range in the vicinity of 11 T is needed to confirm the latter. The NMR spectra recorded in the modulated phase are consistent with a real-space, single-Q modulation of the order parameter, albeit with substantial broadening.

The authors express their appreciation to Drs. Reizo Kato and H. Yamamoto for their contribution of some of the $^{13}$C spin-labelled ET molecules to this project, and to Steve Kivelson for helpful discussions. The work at UCLA was supported by the National Science Foundation, under Grant DMR-1410343. J.W. and H.K. acknowledge the support of the HLD at HZDR, a member of the European Magnetic Field Laboratory (EMFL).

\bibliographystyle{apsrev4-1}

\pagebreak
\onecolumngrid

\begin{center}
\textbf{Supplementary information: Microscopic study of the Fulde-Ferrell-Larkin-Ovchinnikov state in an all-organic superconductor}
\end{center}
\setcounter{equation}{0}
\setcounter{figure}{0}
\setcounter{table}{0}
\setcounter{page}{1}
\makeatletter

\section*{Lineshape under rotation at 10.5 T}

\begin{wrapfigure}[25]{l}{0.5\textwidth}
\begin{center}
\includegraphics[width=0.48\textwidth]{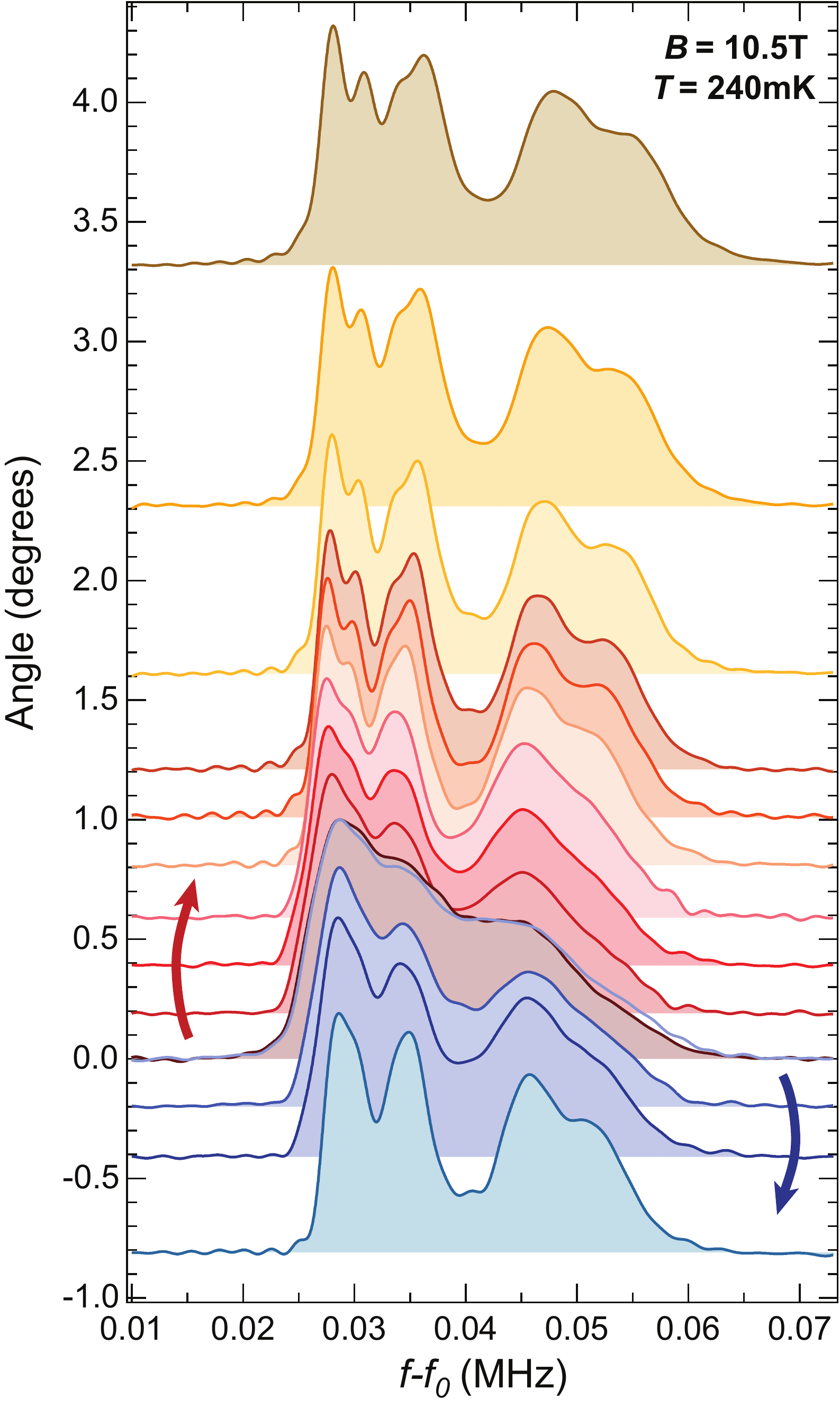}
\caption{Destruction of FFLO/restoration of normal state with rotation about $b$-axis.}
 \label{fig:Rot}
\end{center}
\end{wrapfigure}

The FFLO state is expected to be readily destroyed by rotation of the applied field away from the conducting layers, as demonstrated in the (ET) family of compounds \cite{Beyer:S2013}. In order to verify this tendency, the $^{13}$C NMR spectrum was recorded for small misalignment from the in-plane condition, discussed in the main text, while in the FFLO state ($T$=240 mK, $B$=10.5 T). A piezo-driven rotator allows for single-axis sample rotation with millidegree accuracy. The results are shown in Fig. \ref{fig:Rot}. Misalignment along one direction (red arrow/spectra) as small as 0.2$^{\circ}$ alters dramatically the NMR line shape, and at $\sim 1^{\circ}$ the normal state spectrum is recovered. Returning to the in-plane condition and rotating the opposite direction (blue arrow/spectra) reproduces the same behavior. Thus, the NMR spectral evolution upon small rotations in the FFLO state corroborates the results of specific heat measurements, which suggested a highly sensitive upon sample alignment FFLO phase \cite{Beyer:S2012}. Also, this finding attests to the very precise sample alignment for the measurements described in the main text of this manuscript.
\newpage
\section*{Spin-lattice relaxation recovery profiles}

\begin{figure}[htb]
\begin{center}
\includegraphics[width=3.5in]{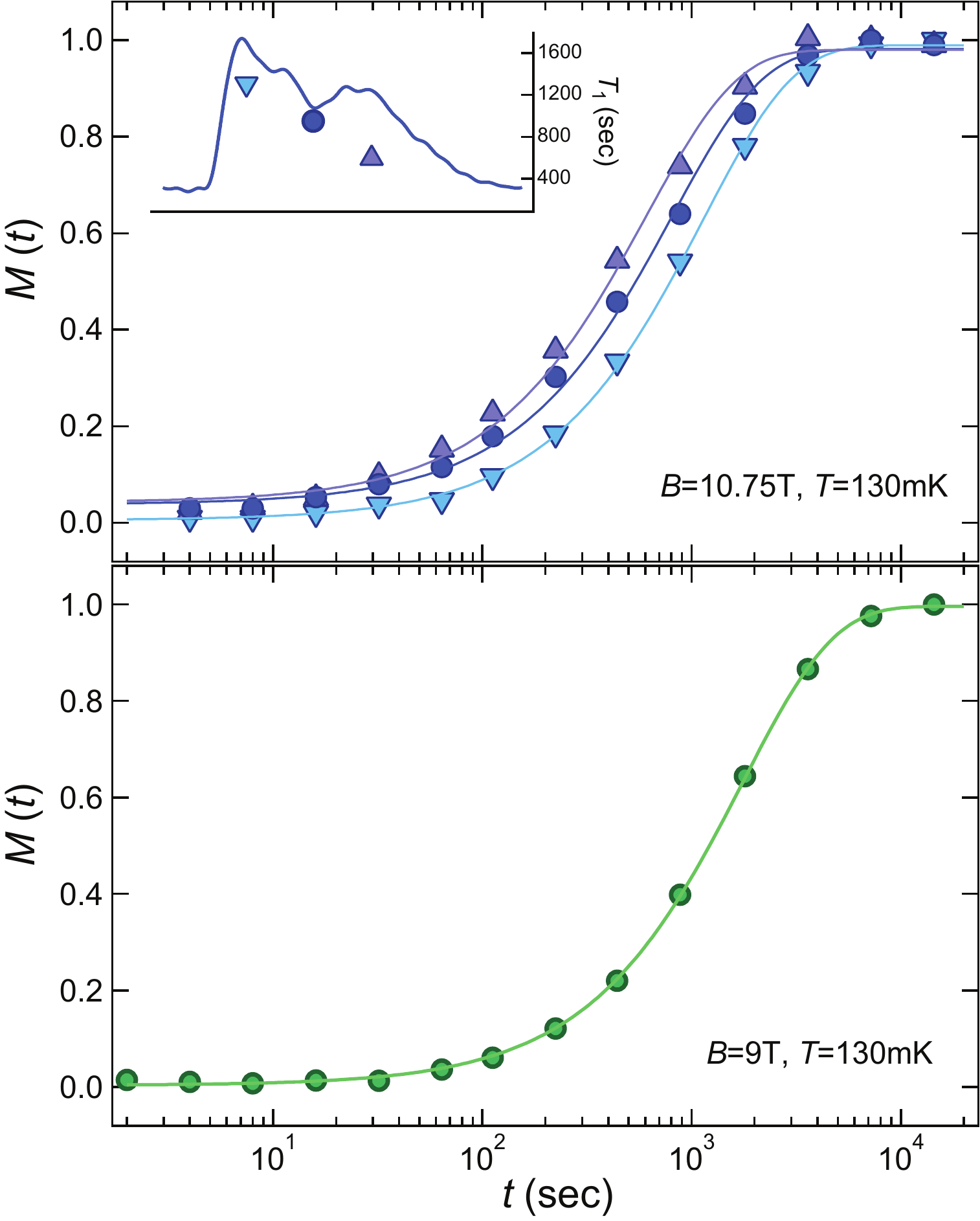}
\caption{Representative $^{13}$C nuclear magnetization recovery profiles after saturation preparation in the uSC (\textit{bottom}) and FFLO (\textit{top}) states. In the uSC state, a single relaxation time characterizes the NMR spectrum, while in the FFLO state there is variation of the relaxation rate over the entire linewidth, as illustrated in the inset. The solid curves are fit to single-exponential recovery $M(t)=M_0(1-e^{-t/T_1})$.}
 \label{fig:RecProf}
 \end{center}
\end{figure}

Figure \ref{fig:RecProf} shows representative nuclear magnetization recovery curves at low-temperature ($T=130$ mK) and for applied field values corresponding to the uSC (\textit{bottom}) and FFLO (\textit{top}) states. In the uSC state at $B=9$ T, the contributions from the inequivalent sites collapse effectively to a single NMR line due to the vanishing spin susceptibility, resulting in a well-defined single-exponential recovery. However, in the FFLO state ($B=10.75$ T), the emergence of a modulated spin polarization, with maxima at the real-space nodes of the order parameter, leads to different hyperfine field for the four inequivalent sites, scaling with the strength of the hyperfine coupling. As a consequence, the spin-lattice relaxation rate, $T_1^{-1}$, also varies across the spectrum, with frequencies associated with larger hyperfine field having faster relaxation rate. In Fig. \ref{fig:RecProf} (\textit{top}), the magnetization recovery curves for the high- and low-frequency end of the NMR spectrum are shown, as denoted in the inset, which describe relaxation rate values differing by approximately a factor of two. The recovery curve corresponding to the entire linewidth is also depicted (circles). The characteristic time scale associated with 63\% of the full recovery of the latter is found to be very close to the mean relaxation time  $\overline{T_1^{-1}(\nu)}$, and it is what is reported in Fig. 3b of the manuscript.

\begin{figure}[ht]
\includegraphics[width=3.5in]{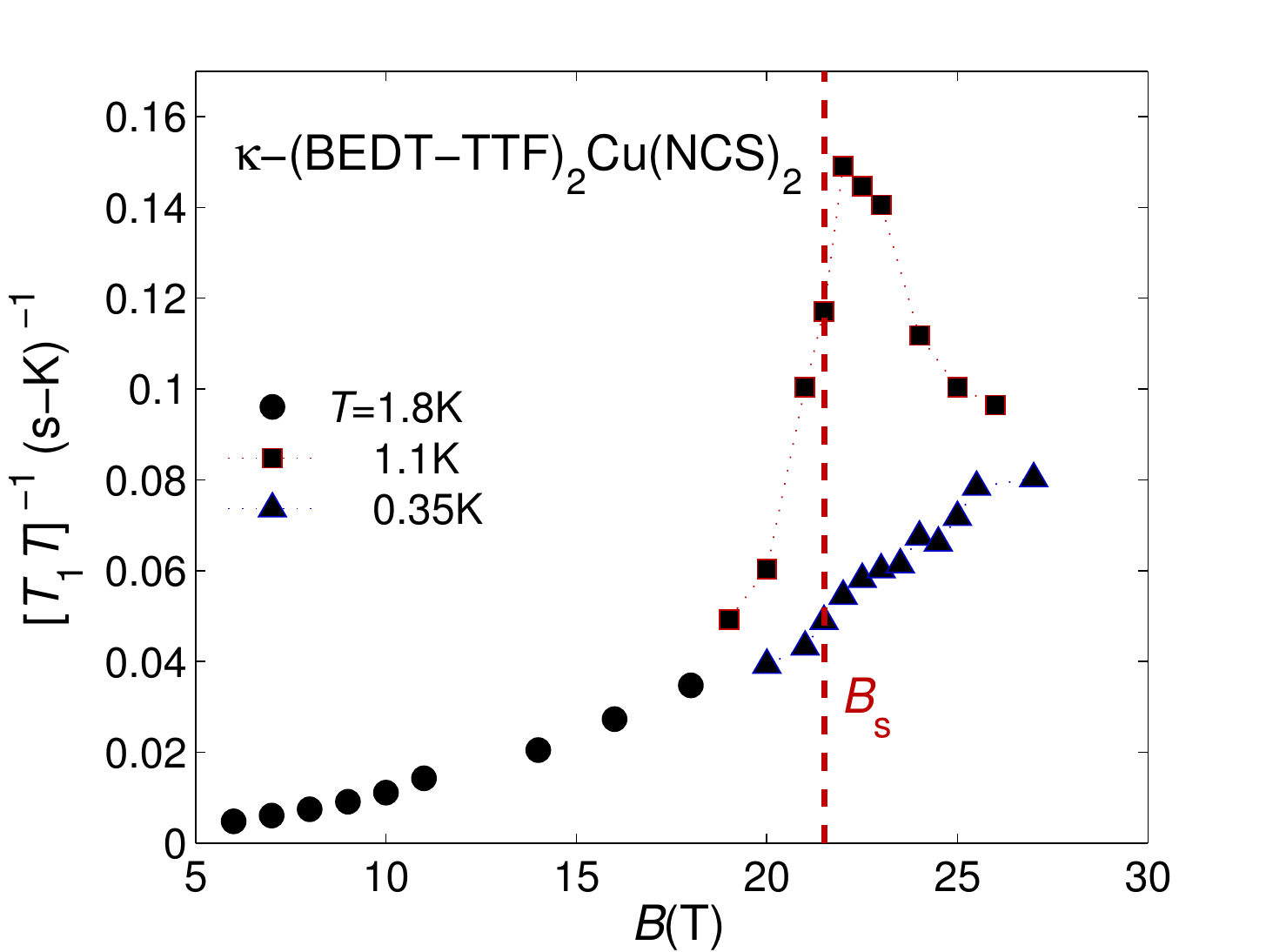}
\caption{$\left(T_1T\right)^{-1}$ as a function of field in $\kappa$-(ET)$_2$Cu(NCS)$_2$. The observed rate enhancement in the FFLO state for $T$=1.1K is ``quenched'' at the lower temperature of $T$=0.35K.}
 \label{fig:kappa}
\end{figure}

It is useful to contrast our relaxation measurements results for the $\beta^{\prime\prime}$-material reported here, and presented in Fig. 3b of the manuscript, to similar measurements on $\kappa$-(ET)$_2$Cu(NCS)$_2$, which were recorded for the work reported in Ref. \cite{Wright:S2011}. Fig. \ref{fig:kappa} shows the field evolution of $\left(T_1T\right)^{-1}$ at various temperature values in the latter material. A large enhancement of $\left(T_1T\right)^{-1}$, compared to the normal-state value, is seen upon entering the FFLO state for $T$=1.1 K. This has been attributed to excess low-energy density of states (DOS) due to bound states formed near the nodes of the FFLO order parameter \cite{Mayaffre:S2014}. However, at lower temperature ($T=0.35$ K), any enhancement of $\left(T_1T\right)^{-1}$ is absent in the FFLO phase, and the observed field dependence is very similar to the one measured for the $\beta^{\prime\prime}$ in Fig. 3b of this manuscript. The observations are consistent with the aforementioned interpretation, since at sufficiently low temperature, the bound state DOS energy exceeds $\epsilon\simeq k_B T$ around the Fermi energy $E_F$, and their influence on the relaxation rate is frozen out.

\section*{Comparison of spectra to single Q sinusoidal modulation of the spin polarization}

\begin{figure}[ht!]
\includegraphics[width=3.5in]{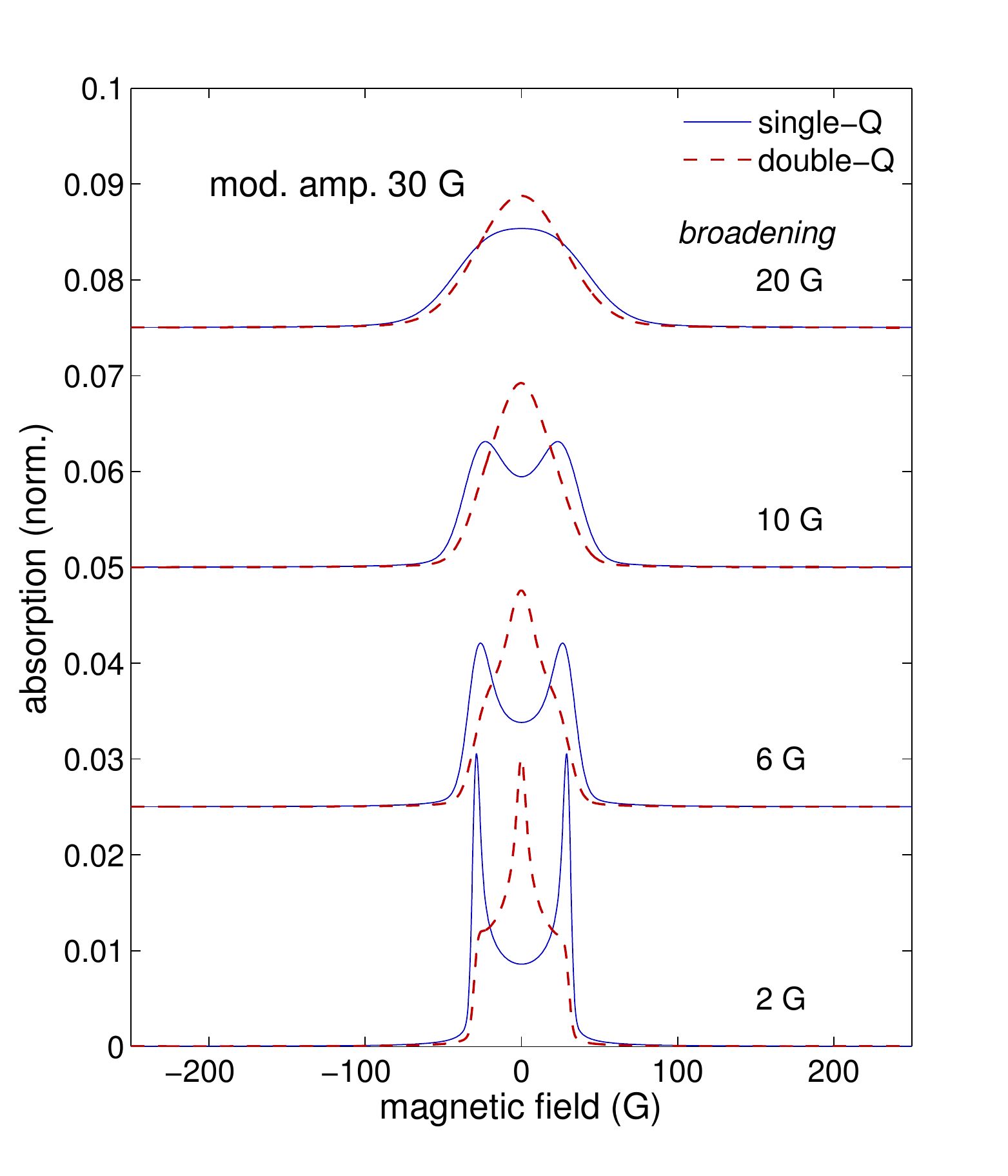}
\caption{Contrast of spectra generated by a single-Q and double-Q sinusoidal modulation of the local field. The total absorption is normalized and the modulation amplitude is set to 30 G. The Gaussian broadening parameter is also given.} \label{fig:singleQdoubleQ}
\end{figure}
NMR is a local probe, so in the incommensurate case that applies to FFLO, the spectra are insensitive to the wavevector $\mathbf{Q}$. On the other hand, the distribution of local fields will depend on whether the modulation is described by a single wavevector, or whether more are required. In Fig. \ref{fig:singleQdoubleQ}, the spectra generated by single-Q and double-Q modulations of 30 G are shown, for what are otherwise equivalent sites. The lineshape is distinct for both cases, provided the Gaussian broadening is small relative to the modulation amplitude. The double-Q case was evaluated for the square lattice. We tentatively associate disorder and subsequent pinning of the wavevector as a source of spectral line broadening.

\begin{figure}
\includegraphics[width=3.5in]{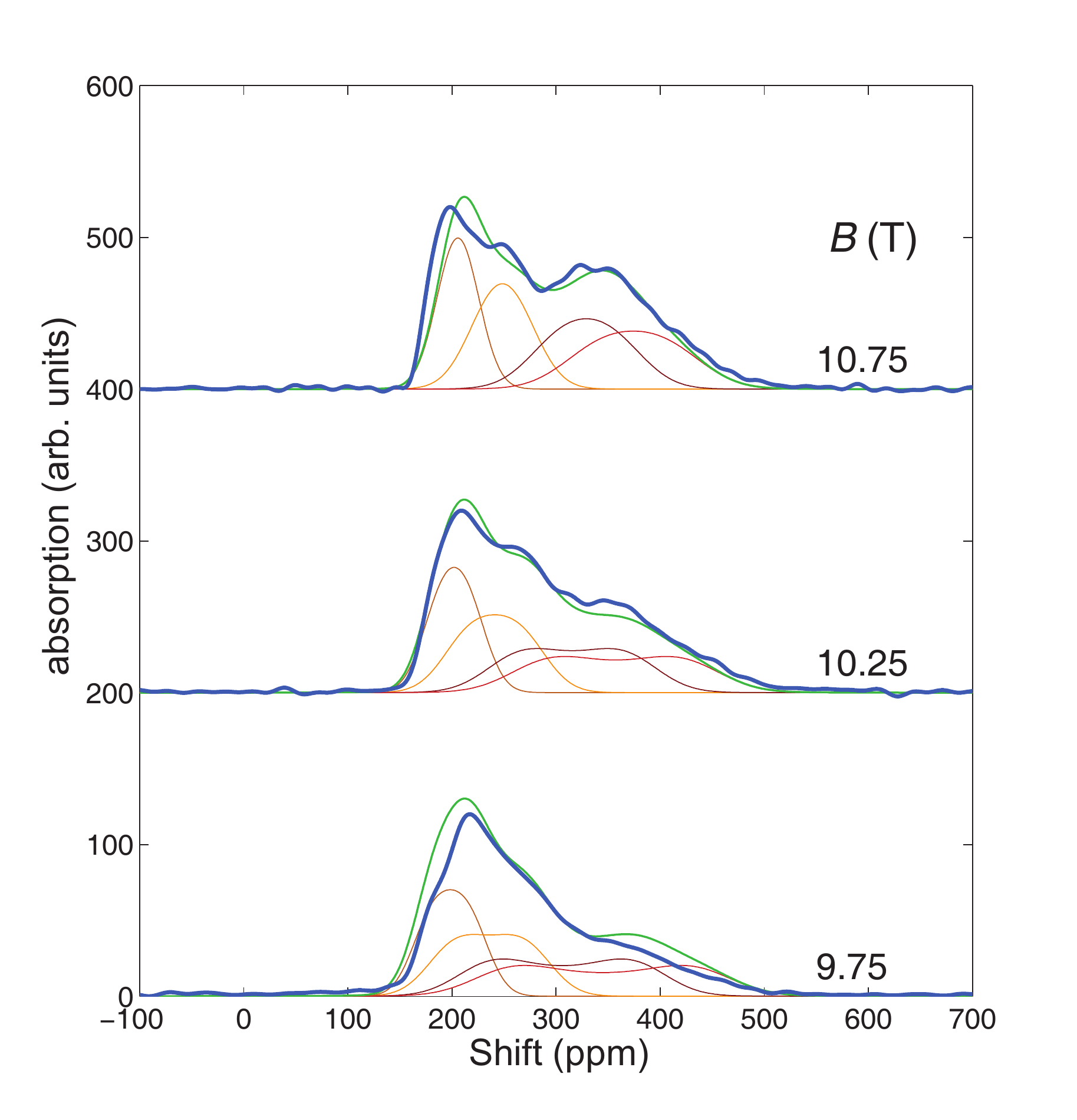}
\caption{Comparison of recorded spectra with that generated by assuming a single-Q sinusoidal modulation of the spin polarization density. Parameters for the model are in Tables \ref{tab:shifts},\ref{tab:Model}}\label{fig:SingleQsimSpectra}
\end{figure}
The experimentally observed lineshape is generated by four sites with inequivalent hyperfine couplings, which presents a challenge because they are not resolved. Thus, we check for consistency with the simplest case, single-Q. The four sites arise from two distinct donor molecule locations, which we have referred to in the main text as \textit{red} and \textit{orange}. Then, on each donor molecule there are spin-labelled central carbon sites, located at crystallographically different locations. The \textit{red} sites are associated with larger hole density and therefore also greater hyperfine fields, leading to greater shifts and faster spin-lattice relaxation. From the collapse of the spectrum deep in the superconducting state (Fig. 2(c), main text), the orbital contribution to the shift is estimated at 150 ppm relative to the standard reference TMS. In comparing to other ET charge-transfer salts, the value is relatively large but not unreasonable. With the orbital shifts approximated, the hyperfine shift in the normal state for each site is determined and included in Table \ref{tab:shifts}. In Fig. \ref{fig:SingleQsimSpectra} are the acquired data for three field values in the FFLO regime, 9.75 T, 10.25 T, and 10.75 T. For each case, the lineshape is modeled by summing lineshapes produced by single-Q sinusoidal modulations from four inequivalent contributions, where constraints are imposed to reduce the number of free parameters. Specifically, the mean shift (\textit{i.e.}, shift of spectral first moment) for each field value is determined from the measured spectrum, from which the average spin polarization relative to the normal state is established (see Tables \ref{tab:shifts},\ref{tab:Model}). The same mean value is applied to each component in the modeling. The hyperfine field modulation amplitude, relative to the normal state shift, is kept identical for each of the four sites. The Gaussian broadening parameter is increasing monotonically with the normal state shift. The implication is that a disorder potential would result in broadened Bragg peaks of the FFLO phase \cite{Cui:S2008}. All parameters are listed in Table \ref{tab:Model}.

\begin{table}[htb]
\begin{tabular}{|c||c|c|}
\hline
abs. peak & $K^{NS}$ & $K_s^{NS}$ \\
\hline\hline
1 & 221 & 71  \\
2 & 275 & 125 \\
3 & 379 & 229 \\
4 & 436 & 286 \\
\hline
\end{tabular}
\caption{$^{13}$C shifts (ppm) for each of the four inequivalent sites, with the field applied in the plane of the molecular layers, $\sim\perp\mathbf{b}$ (as in the manuscript). $K^{NS}$ is the total normal state shift measured relative to TMS, and which includes both hyperfine $K_s^{NS}$ and orbital (chemical) $K_o$ contributions. The chemical shift is estimated at 150 ppm from the low-field and low-temperature spectrum.}
\label{tab:shifts}
\end{table}

\begin{table}[htb]
\begin{tabular}{|c||c|c|}
\hline
B (T) & $\overline{K_s(B)}/K_s^{NS}$ & $\delta h$ \\
\hline\hline
9.75 & 0.67 & 0.36 \\
10.25 & 0.72 & 0.26 \\
10.75 & 0.78 & 0.16 \\
\hline
\end{tabular}
\caption{Mean fractional hyperfine shift $\overline{K_s(B)}/K_s^{NS}$ and modulation amplitude normalized to the normal state shift ($\equiv\delta h$) for each of the recorded spectra appearing in Fig. \ref{fig:SingleQsimSpectra}. Note that the decreasing modulation amplitude is expected for increasing field strength in the FFLO phase. The ratios are kept identical for each of the four sites.}
\label{tab:Model}
\end{table}

\section{The Fermi surface of $\beta''$-(ET)$_2$SF$_5$CH$_2$CF$_2$SO$_3$}

\begin{figure}[htb]
\includegraphics[width=3.5in]{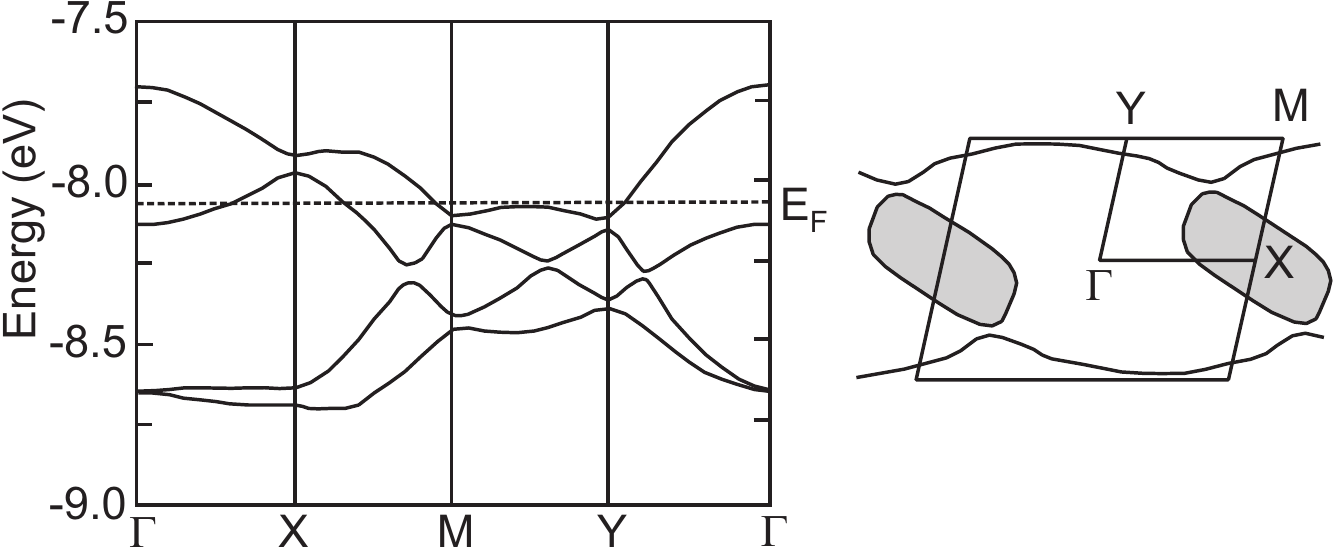}
\caption{Band structure and Fermi surface of $\beta''$-(ET)$_2$SF$_5$CH$_2$CF$_2$SO$_3$, from extended H\"uckel tight-binding calculations.}
 \label{fig:betaDP-FS}
\end{figure}

The quasi-two dimensional electronic structure was calculated using the extended H\"uckel tight-binding method \cite{Geiser:S1996}. Since there are four ET molecules in the unit cell donating two electrons to the counterion sublattice, the Fermi surface occupies an entire Brillouin zone. Zone-folding produces two bands, consisting of a closed pocket and open sheets, which is shown in Fig. \ref{fig:betaDP-FS}. Shubnikov-de Haas measurements reveal beautifully the pocket, although the area inferred is considerably smaller than the tight-binding result. The presence of the open sheets was inferred indirectly from the harmonic content of the oscillations, though direct observation through magnetic breakdown was not achieved \cite{Wosnitza:S2000}.

\bibliographystyle{apsrev4-1}

\begin{thebibliography}{29}
\bibitem [{Dip()}]{DipolarNote}%
  \BibitemOpen
  \href@noop {} {}\bibinfo {note} {Additional unresolved structure is a
  splitting of each of the 4 absorption lines arising from $^{13}$C-$^{13}$C
  dipolar coupling. Its presence is masked here by the relatively greater
  inhomogeneous line broadening.}\BibitemShut {Stop}%
\bibitem [{Ton()}]{ToneNote}%
  \BibitemOpen
  \href@noop {} {}\bibinfo {note} {For the distributions of recoveries
  observed, the distinction from the mean relaxation time
  $\overline{T_1^{-1}(\nu)}$ is small.}\BibitemShut {Stop}%
\bibitem [{Kou()}]{Koutroulakis:2015}%
  \BibitemOpen
  \href@noop {} {}\bibinfo {note} {Other indications for a nodal order
  parameter include $T_1^{-1}\sim T^3$ at low fields, and absence of
  Hebel-Slichter coherence peak for $T<T_c$, to be presented
  elsewhere.}\BibitemShut {Stop}%
\end{thebibliography}

\end{document}